\documentstyle[12pt]{article}

\newtheorem{cor}{Corollary}
\newtheorem{theo}{Theorem}

\def\ba{\begin{eqnarray}}
\def\ea{\end{eqnarray}}
\def\be{\begin{equation}}
\def\ee{\end{equation}}

\newfont{\msbm}{msbm10}
\newfont{\msbms}{msbm6}  
\newfont{\cmss}{cmss10}  

\def\cst{$C^{\ast}$}

\def\U{{\cal U}}
\def\V{{\cal V}}

\def\A{{\cal A}}

\def\E{{\cal E}}

\def\H{{\cal H}}

\def\G{{\cal G}}

\def\eia{\E_{\infty}^{\ast}}

\def\Ep{\epsilon}

\def\ag{{{\cal A}/{\cal G}}}
\def\agb{{\overline \ag}}

\def\Lagmr{L^2\bigl(\agb,\m_r\bigr)}
\def\Lagm0{L^2\bigl(\agb,\m_0\bigr)}

\def\Leiamd{L^2\bigl(\eia,\m_{\Delta}\bigr)}

\def\Dif{Dif\!f}

\def\m{\mu}

\def\f{\phi}

\def\vf{\varphi}

\def\a{\alpha}
\def\b{\beta}

\def\d{\delta}
\def\l{\lambda}

\def\h2{${\rm h}(2)$}

\def\R{\hbox{\msbm R}}
\def\C{\hbox{\msbm C}}

\begin{document}

\baselineskip=18pt


\title{
Invariance properties of induced Fock measures for U(1) holonomies}

\author{J. M. Velhinho}


\date{{\it Departamento de F\'\i sica \\Universidade da Beira 
Interior
\\R. Marqu\^es D'\'Avila e Bolama\\
6201-001 Covilh\~a, Portugal\\jvelhi@mercury.ubi.pt
}}

\maketitle

\begin{abstract}

\noindent  
We study invariance properties of
the measures in the space of generalized U(1) connections
associated to Varadarajan's r-Fock representations.

\vskip .5 in
\centerline{MSC: 83C47, 58B99} 
\end{abstract}

\newpage 

\pagestyle{myheadings}


\section{Introduction}
\label{int}

Holonomies are the starting point for a rigorous approach to quantum gravity -- often called "loop quantum gravity"  --  carried throughout the last decade. 
It is based on Ashtekar's formulation of general relativity as a gauge theory
\cite{As}, loop variables \cite{GT,RS}, \cst -algebra techniques \cite{AI,B1} and integral and functional calculus in spaces of generalized 
connections~\cite{AL1,AL3,MM,B2} 
(an excelent review of both the fundamentals and 
the most recent developments in this field can be found in~\cite{T}). Since the early days of this approach, free 
Maxwell theory  has been a preferred testing ground for new ideas, especially in what concerns the relation between background independent representations of holonomy algebras  and the standard Fock representation for smeared 
fields~\cite{ARS,AI1,AR}. Recently, Varadarajan 
revisited this subject, and proposed a family of representations for 
a kinematical Poisson algebra of $U(1)$ holonomies and certain 
functions of the
electric fields~\cite{Va1,Va2}. Varadarajan's 
work allowed the emergence of 
Fock states within
the framework of generalized connections and 
is therefore a promising starting point to close the gap between 
non-perturbative 
loop quantum gravity states and low energy states~\cite{AL4} (see 
also~\cite{T} for a general discussion of the issue of semiclassical analysis
in loop quantum gravity).

In the present  work we study (quasi-)invariance and mutual singularity
properties of the measures associated to Varadarajan's representations.
These are measures 
on the space $\agb$ of generalized $U(1)$ connections
that can be obtained, by push-forward, from the standard 
Maxwell-Fock measure (see~\cite{AI,AR,ARS} for previous work along these lines
and also~\cite{AL4} for a projective construction of the measures).
We will 
show that the measures  are singular with respect to each other  and are
singular with respect 
to the  measure $\mu_0$ of Ashtekar and Lewandowski.
This implies, in particular, that  the Fock states are not in 
$L^2(\agb,\mu_0)$ (but rather in an extension thereof~\cite{Va2}).
It also follows that
the measures are not quasi-invariant  with respect to the natural action of 
$\agb$ on itself, which is an obstruction to the quantization of the usual
smeared electric fields. On the other hand,
the measures on $\agb$
inherit quasi-invariance properties  directly related to the
electric operators considered by Varadarajan.

This work is organized as follows. In section~\ref{fock} we review 
the Fock representation, whereas 
the loop approach is reviewed in section~\ref{hhm}. 
Varadarajan's measures are presented in section~\ref{ifrep} and
studied in section~\ref{pim}, which contains our main results.
We conclude with a brief discussion, in section~\ref{dis}.

\section{Smeared fields and Fock representation}
\label{fock}
This section briefly reviews some aspects of the Sch\"odinger representation
of the usual Fock space for the Maxwell field, 
following~\cite{GV,RS2,GJ,BSZ}.
We use spatial coordinates $(x^a),\ a=1,2,3$, and units such that 
$c=\hbar=1$.
The Euclidean metric $\delta_{ab}$ in $\R^3$ is used to raise and lower indices whenever necessary. 

As is well known, connections $A$ and electric  fields $E$ do not 
give rise to well defined quantum operators. In the Fock framework, they
are replaced by smeared versions $A(\Ep)=\int A_a\Ep^a d^3x$ and 
$E(\l)=\int\l_aE^ad^3x$, where $\Ep$ belongs to the (nuclear) space 
$\E_{\infty}$ of smooth and fast decaying transverse vector fields
and $\l$ to the (nuclear) space  $(\ag)_{\infty}$ of smooth and fast decaying 
transverse connections. 
The Poisson bracket between these basic observables is
\be
\label{f4}
\bigl\{A(\Ep),E(\l)\bigr\}=\int\l_a\Ep^a d^3x\,,
\ee
to which  correspond the Weyl relations: 
\be
\label{f5}
\V(\l)\,\U(\Ep)=e^{i\int \l_a\Ep^ad^3x}\U(\Ep)\V(\l).
\ee
The usual Fock representation can be realized in the
Hilbert space $\Leiamd$, where $\E_{\infty}^{\ast}$ is the space  of tempered
distributional 1-forms (the topological dual of $\E_{\infty}$) and $\m_{\Delta}$ is the Gaussian measure defined by:
\be
\label{f10}
\int_{\E_{\infty}^{\ast}} e^{i\f(\Ep)}d\m_{\Delta}(\f)  :=  
\exp\left(-{1\over 4}
\int \d_{ab}\Ep^a(-\Delta)^{-1/2}\Ep^bd^3x\right) \,,
\ee
where $\Delta$ is the Laplacian in $\R^3$. 
It is well known that $\m_{\Delta}$ is  quasi-invariant with respect to the 
action of $(\ag)_{\infty}$ on $\E_{\infty}^{\ast}$:
\be
\label{f6}
\E_{\infty}^{\ast}\ni\f\mapsto\f +\l \ \ ,\ \l\in (\ag)_{\infty}\,,
\ee
where $\f +\l\in\E_{\infty}^{\ast}$ is defined by
\be
\label{f7}
(\f +\l)(\Ep):=\f(\Ep)+\int\l_a\Ep^a d^3x\,, \ \ \forall\Ep\in\E_{\infty}\,.
\ee
(Recall that a measure $\mu$ is quasi-invariant with respect to a 
group of transformations $T$ if the push-forward measure $T_{\ast}\mu$ has 
the same zero measure sets as
$\mu$, $\forall T$, i.e.~$(T_{\ast}\mu)(B)=0$ if and only if $\mu(B)=0$.)
One therefore has an unitary representation $\V$ of the abelian group 
$(\ag)_{\infty}$ as translations:
\be
\label{f8}
\bigl(\V(\l)\psi\bigr)(\f)=\sqrt{d\m_{\Delta,\l}(\f)\over d\m_{\Delta}(\f)}\,
\psi(\f-\l)\,,
\ \ \psi\in\Leiamd,
\ee
where $\m_{\Delta,\l}$ is the push-forward of the measure $\m_{\Delta}$ by the 
map (\ref{f6})
and $d\m_{\Delta,\l}/d\m_{\Delta}$ is the Radon-Nikodym
derivative. (The existence of both the Radon-Nikodym derivative 
and of its inverse is  equivalent to quasi-invariance.) 

A representation of the Weyl relations (\ref{f5}) is achieved with the following
representation $\U$ of $\E_{\infty}$:
\be
\label{f9}
\bigl(\U(\Ep)\psi\bigr)(\f)=e^{-i\f(\Ep)}\psi(\f)\,.
\ee
Since both representations $\U$ and $\V$ are continuous, the quantized fields
$\hat A(\Ep)$ and $\hat E(\l)$ 
can be identified with the generators of the one-parameter groups $\U(t\Ep)$ 
and $\V(t\l)$, respectively.

\section{Holonomies and Haar measure}
\label{hhm}

This section briefly reviews the  loop approach to 
the Maxwell field, following in essence the  general framework 
for gauge theories with a compact (not necessarily abelian)  
group (see e.g.~\cite{T} and references therein). 
Notice, however, that the presentation of the uniform measure 
$\mu_0$~\cite{AL1}  and the quantization of 
electric fields~\cite{AL3} are considerably simpler in
the $U(1)$ case. 

In the loop approach the configuration variables are (traces of) holonomies 
rather then smeared connections. Let us then consider  $U(1)$ holonomies
\be
\label{f11}
T_{\a}(A):=e^{i\oint_{\a}A_adx^a}
\ee
associated with piecewise analytic loops on $\R^3$. 
It is convenient to eliminate redundant loops, i.e.~one identifies two
loops $\a$ and $\b$ 
such that $T_{\a}(A)=T_{\b}(A)\ \forall A$.
Such classes of loops are called hoops. The set $\H\G$ of all 
$U(1)$ hoops is an abelian group under the natural composition 
of loops. 

The set of holonomy functions $T_{\a}$, ${\a}\in \H\G$, is an abelian
$\ast$-algebra. The \cst\ completion in the supremum norm is  called the $U(1)$ 
holonomy algebra $\overline{\H\A}$~\cite{AI,AL1}. It turns out that 
$\overline{\H\A}$ is isomorphic to the algebra of continuous functions on the
space $\agb$ of generalized connections, where $\agb$ is the set  of all group 
morphisms from the
hoop group $\H\G$ to $U(1)$.
In order to describe the isomorphism, let us consider the
functions 
$\Psi_{\a}:\agb \to U(1)$:
\be
\label{f13}
\bar A\mapsto\Psi_{\a}(\bar A):= \bar A(\a)\,,\ 
\ee
where 
$\a\in \H\G$ and $\bar A(\a)$ denotes evaluation.
The space $\agb$ is  compact 
in the weakest topology such 
that all
functions $\Psi_{\a}$ are continuous. It is a key result that $\agb$ is 
homeomorphic to the spectrum of 
$\overline{\H\A}$~\cite{MM,AL1,AL2},
with the functions $\Psi_{\a}$ 
corresponding to 
$T_{\a}$.

(Cyclic) representations of $\overline{\H\A}$ 
are in 1-1 correspondence with positive linear functionals on 
$\overline{\H\A}$. By the above isomorphism,
those 
are in turn in 1-1 correspondence with Borel measures in $\agb$.
Given a  measure $\m$, one thus has a representation $\Pi$ of
$\overline{\H\A}$ in the Hilbert space $L^2\bigl(\agb,\m\bigr)$: 
\be
\label{f17}
\bigl(\Pi(T_{\a})\psi\bigr)(\bar A)=\Psi_{\a}(\bar A)\psi(\bar A),\ \ \forall\psi\in L^2\bigl(\agb,\m\bigr)\,.
\ee
The associated positive linear functional $\vf$ is defined by:
\be
\label{vf}
\vf(T_{\a})=\langle 1,\Pi(T_{\a})1\rangle=\int_{\agb}\Psi_{\a}\, d\m\,.
\ee
In the $U(1)$ case,  $\agb$ is a topological 
group~\cite{AL1,Ma} with multiplication
\be
\label{f14}
\bigl({\bar A}'\bar A\bigr)(\a)={\bar A}'(\a)\bar A(\a),\ \ {\bar A}',\bar A\in \agb,\ \a\in\H\G,
\ee
and inverse
${\bar A}^{-1}(\a)=\bar A(\a^{-1}).$
Let us consider the Haar measure $\m_0$ and the associated 
representation $\Pi_0$ of $\overline{\H\A}$~\cite{AL1}.
Since $\m_0$ is invariant, we also have an unitary representation 
$V_0$ of the group $\agb$ in $L^2\bigl(\agb,\m_0\bigr)$:
\be
\label{f18}
\bigl(V_0({\bar A}')\psi\bigr)(\bar A)=\psi({\bar A}'\bar A),\ \ \forall\psi\in L^2\bigl(\agb,\m_0\bigr)\,.
\ee
The representation $V_0$ leads to smeared electric operators, as follows.
For $\l\in(\ag)_{\infty}$,
let $\bar A_{\l}$ denote the element of $\agb$ defined 
by holonomies, i.e.~$\bar A_{\l}(\a):=T_{\a}(\l),\ \forall\a\in\H\G$.
Restricting $V_0$ to  elements $\bar A_{\l}$ 
and the representation $\Pi_0$ to the functions $T_{\a}$, one obtains 
the commutation relations:
\be
\label{f20}
V_0(\l)\Pi_0(T_{\a})=e^{i\oint_{\a}\l_adx^a}\Pi_0(T_\a)V_0(\l)\,,
\ee
where $V_0(\l):=V_0(\bar A_{\l})$. 
The action of $V_0(\l)$ is particularly simple for the dense space
of finite linear combinations
of functions $\Psi_{\a}$:
\be
\label{new2}
V_0(\l)\Psi_{\a}=e^{i\oint_{\a}\l_adx^a}\Psi_{\a}\, .
\ee
Let us consider the  one-parameter unitary group $V_0(t\l)$, $t\in\R$,
and let $dV_0(\l)$ be its self-adjoint generator.
>From (\ref{f20}) one finds  the  
commutator:
\be
\label{f21}
\bigl[dV_0(\l),\Pi_0(T_{\a})\bigr]=\left(\oint_{\a}
\l_adx^a\right)\Pi_0(T_{\a})\,,
\ee
showing that the operators $\Pi_0(T_{\a})$ and $dV_0(\l)$ give a quantization
of the Poisson algebra of holonomies $T_{\a}$ and smeared electric fields
$E(\l)$, in $\Lagm0$. In this representation, the states
$\Psi_{\a}$ describe one-dimensional excitations of the electric 
field along loops, or electric flux "quanta", and are therefore called loop
states~\cite{GT,RS}. These type of excitations are, of course, absent in Fock
space. On the other hand, nor are the familiar Fock $n$-particle states or 
coherent states obviously related to loop states.

\section{$r$-Fock measures}
\label{ifrep}

In this section we present Varadarajan's
$r$-Fock representations of the $U(1)$ holonomy algebra 
$\overline {\H\A}$ from the 
measure theoretic point of view.

Let us start with hoop form factors~\cite{ARS,AR,AI}.
Given a hoop $\a$, the form factor $X_{\a}$ is  the 
transverse distributional vector field such that
\be
\label{f23}
\int X_{\a}^a(x)A_a(x)d^3x=\oint_{\a}A_adx^a\,,\ \ \ \forall A.
\ee
Consider the one-parameter family of functions on $\R^3$: 
\be
\label{f25}
f_r(x)={1 \over 2\pi^{3/2}r^3}\,e^{-x^2/2r^2}\,,
\ee
where $r>0$. The smeared form factors are smooth and fast decaying transverse 
vector fields, 
i.e.~elements of $\E_{\infty}$, defined by:
\be
\label{f26}
X_{\a,r}^a(x):=\int f_r(y-x)X_{\a}^a(y)d^3y\,.
\ee
One thus has, for each  
$r$, a map $\a\mapsto X_{\a,r}$  from hoops to  $\E_{\infty}$. 
Notice that  the composition of hoops is preserved, 
i.e.~$X_{\a\b,r}=X_{\a,r}+X_{\b,r}$~\cite{AR}.

Smeared form factors can be used to define measurable maps from 
the space
of distributional connections
$\E_{\infty}^{\ast}$
to  $\agb$. 
Consider then the family of maps 
$\Theta_r: \E_{\infty}^{\ast}\rightarrow\agb$ given by 
$\f\mapsto \bar A_{\f,r}$, where
\be
\label{f27}
\bar A_{\f,r}(\a):=e^{i\f(X_{\a,r})}
\ \ \ \forall\a\in\H\G.
\ee
Since the $\sigma$-algebra of measurable sets in $\agb$ is the smallest one 
such that all 
functions $\Psi_{\a}$ (\ref{f13}) are measurable, one sees that 
$\Theta_r$ is measurable if and only if the maps 
$\Psi_{\a}\circ\Theta_r:\E_{\infty}^{\ast}\rightarrow U(1)$ 
are measurable for all $\a\in\H\G$, 
which is  clearly true, since they can be obtained as a composition of 
measurable maps: 
$\f\mapsto\f(X_{\a,r})\mapsto e^{i\f(X_{\a,r})}$.

One can now use the maps $\Theta_r$ to push-forward the Fock measure 
$\m_{\Delta}$, thus obtaining a family of 
measures $\m_r:=(\Theta_r)_{\ast}\m_{\Delta}$ on $\agb$.
By definition
\be
\label{f28}
\m_r(B)=\m_{\Delta}(\Theta_r^{-1}B)\ \ \ 
\forall\ {\rm measurable\ set}
\ B\subset\agb.
\ee
Each of the measures $\m_r$
provides us with a Hilbert space $L^2\bigl(\agb,\m_r\bigr)$ and a 
representation $\Pi_r$ of  
$\overline{\H\A}$. 
The associated positive linear functional $\vf_r$
is:
\be
\label{f30}
\vf_r(T_{\a})=\int_{\agb}\Psi_{\a}(\bar A)\, d\m_r(\bar A)=\int_{\E_{\infty}^{\ast}}
e^{i\f(X_{\a,r})}d\m_{\Delta}(\f)\, .
\ee
Expression (\ref{f30}) shows that the representation $\Pi_r$ is the 
$r$-Fock representation considered by Varadarajan in \cite{Va1}, 
$\m_r$ being the $r$-Fock measure in $\agb$ whose existence was proved 
in \cite{Va2}.

\section{Properties of the $r$-Fock measures}
\label{pim}
The present section contains our main results. We show that the $r$-Fock 
measures
$\m_r$ are all mutually singular, and are singular with respect to the Haar
measure $\m_0$. We study also (quasi-)invariance properties of the $r$-Fock 
measures  $\m_r$ 
and their relation to the quantization of certain twice smeared electric fields
introduced in \cite{Va1}.

Let $\Dif$\/ be the group of (analytic) diffeomorphisms of $\R^3$. The natural action of $\Dif$ on the (piecewise analytic) curves of $\R^3$ induces an action on the hoop group  $\H\G$:
\be
\label{f34}
\H\G\times\Dif\ni (\a,\vf)\mapsto \vf\a\,,
\ee
and therefore one has an action of $\Dif$ in $\agb$, given by
\be
\label{f35}
\bigl(\vf^{\ast}\bar A\bigr)(\a)=\bar A(\vf\a),\ \ \vf\in\Dif,\ \bar A\in\agb,\ \a\in\H\G\,.
\ee
It can be seen that the maps $\vf^{\ast}:\agb\rightarrow\agb$ are 
continuous \cite{AL1,AL2,B1}. The Haar measure $\m_0$ is invariant 
under the action of $\Dif$, since no background geometrical structure 
is used in its definition \cite{AL1}. The induced measures $\m_r$, 
on the other hand, are  not invariant, due to the appearance of the Euclidean metric $\d_{ab}$ in the construction of the Fock measure $\m_{\Delta}$. 
>From now on we will restrict our attention to the Euclidean group, i.e., the subgroup of $\Dif$ of transformations that preserve the Euclidean metric. It is clear that the measures $\m_r$ are invariant under these transformations, given the well known Euclidean invariance of the Fock measure.

Besides being invariant, the Fock measure is moreover ergodic with respect 
to the action of the Euclidean group (see e.g.~\cite{BSZ,Ve2}), 
which means that the only invariant functions in $\Leiamd$  are the 
constant functions.
This ergodic property is shared by the  measures $\m_r$, since if an 
invariant and non-constant function $\psi$ were to exist in 
$\Lagmr$, then the pull-back $\psi\circ\Theta_r$ would define an 
invariant and non-constant function in $\Leiamd$.

An important fact is that the Haar measure $\m_0$ on $\agb$ is also ergodic 
under the 
action of the Euclidean group, as follows from more general results proven 
in \cite{MTV}. Thus, all measures $\m_r,\ r\in\R^+$, and $\m_0$ are invariant 
and ergodic 
under the action of the same group. From well known results in measure theory 
(see e.g. \cite{Ya}), this is only possible if all these measures are 
mutually singular, meaning that each measure of the set 
$\{\m_r,\ r\in\R^+\}\cup\m_0$ is supported on a subset of $\agb$ which 
has zero measure with respect to all the other measures (recall that a 
subset $X$ of a space $M$ is said to be a support for the measure 
$\m$ on $M$ if any measurable subset $Y$ on the complement, $Y\subset X^c$, 
has measure zero). It is thus proven that
\begin{theo}
\label{t2}
The measures in the set $\{\m_r,\ r\in\R^+\}\cup\m_0$ are all singular with respect to each other.
\end{theo}
Theorem \ref{t2} leads to the conclusion that none of the measures $\m_r$ is 
quasi-invariant under the action of $\agb$ on itself. This follows from 
the fact that $\agb$ is a compact group, which implies that any 
quasi-invariant measure is in the equivalence class of the Haar measure, 
meaning that it must have the same zero measure sets (see e.g. \cite[9.1]{Ki}). Thus
\begin{cor}
\label{c1}
The measures $\m_r,\ r\in\R^+$, are not quasi-invariant.
\end{cor}
We saw in section \ref{hhm} how the quantization of smeared electric fields
can be obtained from an unitary representation of the group $\agb$ in the Hilbert space 
$\Lagm0$. From  the corollary we conclude that such an unitary representation of $\agb$ is not available in the Hilbert spaces $\Lagmr$. One should thus look for the quantization of different functions of the electric fields.

Varadarajan showed in \cite{Va1} that certain "Gaussian-smeared smeared" 
electric fields can be consistently quantized in the $r$-Fock representations. 
In the remaining we will relate the quantization of these functions
to quasi-invariance properties of the $r$-Fock measures $\m_r$.
We will start by establishing the quasi-invariance properties, which, 
as expected,
follow from the quasi-invariance of the Fock measure under the 
action (\ref{f6}).

Let us consider the restriction of the maps $\Theta_r$ (\ref{f27}) to 
$(\ag)_{\infty}$, i.e., we consider the maps
\be
\label{f36}
(\ag)_{\infty}\ni\l\mapsto\bar A_{\l,r}\in\agb
\ee
such that
\be
\label{f37}
\bar A_{\l,r}(\a)=\exp\left(i\int\l_a(x)X_{\a,r}^a(x)d^3x\right),\ \ \forall\a\in\H\G\,.
\ee
It is clear that $\bar A_{\l +\l',r}=\bar A_{\l,r}\bar A_{\l',r}$, and 
therefore the group $(\ag)_{\infty}$  acts on the space $\agb$ as a 
subgroup of the full group $\agb$. Let us denote this action by $\Xi_r$:
\be
\label{f38}
(\ag)_{\infty}\times\agb\ni(\l,\bar A)\mapsto\bar A_{\l,r}\bar A\,.
\ee
For any given $\l\in(\ag)_{\infty}$, let $\m_{\l,r}$ denote the
push-forward of the measure $\mu_r$ by the map 
$\bar A \mapsto \bar A_{\l,r}\bar A$. The measure $\m_{\l,r}$ is completely determined by the integrals of continuous functions $F(\bar A)$:
\be
\label{f39}
\int_{\agb} F(\bar A)d\m_{\l,r}(\bar A)=\int_{\agb} F\bigl(\bar A_{\l,r}\bar A\bigr)d\m_r(\bar A)\,.
\ee
We need only to consider the functions 
$\Psi_{\a}$~(\ref{f13}),  and therefore the measure 
$\m_{\l,r}$ is  determined by the following map from $\H\G$ to $\C$:
\be
\label{f40}
\a\mapsto\int_{\agb}\bigl(\bar A_{\l,r}\bar A\bigr)(\a)d\m_r(\bar A)\,.
\ee
One gets from (\ref{f14}), (\ref{f30}) and (\ref{f37}):
\be
\label{f41}
\!\!\!\!\!\!\int_{\agb}\bigl(\bar A_{\l,r}\bar A\bigr)(\a)d\m_r(\bar A)  = \exp\left(i\int\l_a(x)X_{\a,r}^a(x)d^3x\right)\int_{\eia}e^{i\f(X_{\a,r})}d\m_{\Delta}(\f).
\ee
Recalling the action (\ref{f6}) of $(\ag)_{\infty}$ on 
$\E_{\infty}^\ast$, one gets further:
\begin{eqnarray}
\int_{\agb}\bigl(\bar A_{\l,r}\bar A\bigr)(\a)d\m_r(\bar A) & = &
\int_{\eia}e^{i(\f+\l)(X_{\a,r})}d\m_{\Delta}(\f) \nonumber \\
\label{f42}
& = & \int_{\eia}e^{i\f(X_{\a,r})}d\m_{\Delta,\l}(\f)\,,
\end{eqnarray}
where the measure $\m_{\Delta,\l}$  is the push-forward of 
$\m_{\Delta}$ by the map $\f\mapsto\f+\l$  (\ref{f6}).  
Recalling also the arguments of section \ref{ifrep},
one  sees easily that the measure $\m_{\l,r}$ coincides with $(\Theta_r)_{\ast}\m_{\Delta,\l}$, the push-forward of $\m_{\Delta,\l}$ by the map $\Theta_r$ (\ref{f27}). Since the Fock measure $\m_{\Delta}$ is quasi-invariant under the action of $(\ag)_{\infty}$, this is sufficient to prove that, for any $r\in\R^+$, the measure $\m_r$ is quasi-invariant with respect to the action 
$\Xi_r$ (\ref{f38}). For if $B\subset\agb$ is such that $\m_r(B)=0$ we then 
have $\m_{\Delta}(\Theta_r^{-1}B)=0$, by definition of the push-forward 
measure. The quasi-invariance 
of $\m_{\Delta}$ then shows that 
$\m_{\Delta,\l}(\Theta_r^{-1}B)=0$, $\forall\l\in (\ag)_{\infty}$, 
which in turn is equivalent to $\m_{\l,r}(B)=0$. Thus
\begin{theo}
\label{t3}
The measure $\m_r$ is quasi-invariant with respect to the action $\Xi_r$, 
for any given $r$.
\end{theo}
Using this result, we define, for any $r$, a natural unitary representation $V_r$ of $(\ag)_{\infty}$ in $\Lagmr$:
\be
\label{f43}
\bigl(V_r(\l)\psi\bigr)(\bar A)=\sqrt{d\m_{\l,r}\over d\m_r}\,\psi
\bigl(\bar A_{\l,r}\bar A\bigr),\ \ \l\in (\ag)_{\infty}, \ \psi\in\Lagmr\,,
\ee
where $d\m_{\l,r}/d\m_r$ is the Radon-Nikodym derivative.

One can  easily work out the commutation relations 
between $V_r(\l)$ and the $r$-Fock representation $\Pi_r(T_{\a})$ of holonomies:
\be
\label{f44}
V_r(\l)\Pi_r(T_{\a})=\exp\left(i\int\l_a(x)X_{\a,r}^a(x)d^3x\right)
\Pi_r(T_{\a})V_r(\l)\,.
\ee
Let us consider the self-adjoint generator $dV_r(\l)$ of the one-parameter 
unitary group $V_r(t\l)$, $t\in\R$.
Notice that the existence of $dV_r(\l)$, or the continuity
of the one-parameter group $V_r(t\l)$, follows from the continuity
of the representation $\V$~(\ref{f8}) in $\Leiamd$.
>From (\ref{f44}) one obtains the following commutator:
\be
\label{f46}
\bigl[dV_r(\l),\Pi_r(T_{\a})\bigr]=\left(\int\l_a(x)X_{\a,r}^a(x)d^3x\right)\Pi_r(T_{\a})\,.
\ee
This commutator is indeed the quantization of a given classical Poisson 
bracket, as realized by Varadarajan \cite{Va1}. Consider then, for each $r$, 
the following functions of the electric field, parametrized by elements of 
$(\ag)_{\infty}$:
\be
\label{f47}
E^a(x)\mapsto E_r(\l):=\int\l_a(x)\left(\int f_r(x-y)E^a(y)d^3y\right)d^3x\,,
\ee
where $f_r$ is given by (\ref{f25}). The functions $E_r(\l)$ are referred 
to as "Gaussian-smeared smeared electric fields" in \cite{Va1}. 
The Poisson bracket between these functions and the holonomies is
\be
\label{f48}
\bigl\{T_{\a},E_r(\l)\bigr\}=i\left(\int\l_a(x)X_{\a,r}^a(x)d^3x\right)T_{\a}\,,
\ee
showing that $dV_r(\l)$ can be seen as the quantization in $\Lagmr$ of the classical function $E_r(\l)$.

\section{Discussion}
\label{dis}
Varadarajan's use of smeared form factors allowed an embedding of distributional
connections into $\agb$, overcoming the fact that the natural embedding of 
connections is not extensible to distributions. This step is very welcome,
since $\agb$ can be seen as a common measurable space from which both
Fock states and loop states can be defined. 
In particular, the $r$-Fock
measures $\m_r$ give 
natural images $L^2\bigl(\agb,\m_r\bigr)$ 
of the Fock space~\cite{Va1}. 
The Fock states, however, cannot be regarded as elements of 
$\Lagm0$, the kinematical Hilbert space for loop states, as a consequence 
of the mutual singularity
between the Haar measure and the $r$-Fock measures, which we have proven.
Nevertheless,
Fock states can be exhibited within the framework of loop states. 
In fact, Varadarajan~\cite{Va2} showed that the Fock states 
can be realized as elements of a natural extension of $\Lagm0$, 
e.g.~the dual $Cyl^{\ast}$ of the 
space of
cylinder functions in $\agb$. 
Notice that such an extension from 
$\Lagm0$ to $Cyl^{\ast}$ is already required in loop quantum gravity,
in order to solve the
diffeomorphism constraint~\cite{ALMMT}.
A suitable generalization of Varadarajan's work to quantum gravity is, 
therefore,
expected to produce an embedding of Minkowskian Fock-like states
(describing  "gravitons" of a semiclassical or low energy effective theory)
into the space  $Cyl^{\ast}$ of non-perturbative physical loop states.
These issues are currently under investigation (see~\cite{AL4} and 
also~\cite{T} for a more general approach to semiclassical analysis).
In these efforts, measure theory in $\agb$  plays a
relevant role, e.g.~in the definition of quantum operators and in the
analysis of the 
physical contents of the states. A good understanding of the $r$-Fock 
measures and their relation to the Haar measure may therefore be important
to further developments.

In order to complement our measure theoretical results, we
would like to conclude with a brief comment regarding topological aspects.
Although the $r$-Fock measures are supported in irrelevant sets with respect 
to the Haar measure, it can be shown on the topological side
that every conceivable support of a $r$-Fock measure is 
dense in $\agb$~\footnote{The crucial result, pointed out by
M.~Varadarajan, is that $\Theta_r(\E_{\infty}^{\ast})$ is dense.
The denseness of  smaller supports follows from the faithfulness of
the Fock measure $\m_{\Delta}$ in $\E_{\infty}^{\ast}$ and the
continuity of $\Theta_r$  with respect to an appropriate 
topology in  $\E_{\infty}^{\ast}$.}. The $r$-Fock measures $\m_r$ are therefore 
faithful~\footnote{An independent proof using projective arguments 
is given in~\cite{AL4}.}, just like the  
Haar measure $\m_0$~\cite{AL1}.
(Recall that a Borel measure is said to be faithful if every non empty open set
has non zero measure, which is readily seen to be equivalent to the denseness
of every conceivable support. Turning to representations, a measure in $\agb$
is faithful if and only if the corresponding representation of the holonomy
algebra $\overline{\H\A}$ is faithful.) 
The fact that the Haar measure and the $r$-Fock measures are all faithful 
and mutually 
singular means that one can find a family of mutually disjoint dense sets,
each of which supports a different measure. 
Notice finally that dense sets in $\agb$ that do not contribute
to the Haar measure were already known, e.g.~the set of smooth 
connections~\cite{MM} or a considerable extension of it given in~\cite{MTV}. 
In the present 
case, however, one has new measures, living on $\m_0$-irrelevant sets.


\section*{Acknowledgements}
\noindent I thank Jos\'e Mour\~ao,  
Jerzy Lewandowski, Madhavan Varadarajan and Roger Picken.
This work was supported in part by 
PRAXIS/2/2.1/FIS/286/94 and
CERN/P/FIS/40108/2000.

\newpage






\begin{thebibliography}{ALMMT}

\bibitem[As]{As} A. Ashtekar, {\it Lectures on non-Perturbative Canonical
Quantum Gravity} (World Scientific, Singapore, 1991).

\bibitem[ARS]{ARS} A. Ashtekar, C. Rovelli, L. S. Smolin, Phys. Rev. {\bf D44}, 1740 (1991).

\bibitem[AR]{AR} A. Ashtekar, C. Rovelli, Class. Quantum Grav. {\bf 9}, 1121
(1992).



\bibitem[AI1]{AI1} A. Ashtekar, C. J. Isham, Phys. Lett. {\bf B274}, 393 
(1992).


\bibitem[AI2]{AI} A. Ashtekar, C. J. Isham,
Class. Quant. Grav. {\bf 9}, 1433 (1992).



\bibitem[AL1]{AL1} A. Ashtekar, J. Lewandowski, {\it Representation
Theory of Analytic Holonomy $C^\star$ Algebras}, in Knots and 
Quantum Gravity, ed.\ J. Baez (Oxford University Press,
Oxford, 1994).

\bibitem[AL2]{AL2} A. Ashtekar, J. Lewandowski, 
J. Math. Phys. {\bf 36}, 2170 (1995).

\bibitem[AL3]{AL3} A. Ashtekar, J. Lewandowski, 
J. Geom. Phys. {\bf 17}, 191 (1995).

\bibitem[AL4]{AL4} A. Ashtekar, J. Lewandowski, 
Class. Quant. Grav. {\bf 18}, L117 (2001).


\bibitem[ALMMT]{ALMMT} A. Ashtekar, J. Lewandowski, D. Marolf,
J. Mour\~ao, T. Thiemann, J. Math. Phys. {\bf 36}, 6456 (1995).


\bibitem[BSZ]{BSZ}  J. Baez, I. Segal, Z. Zhou, {\it Introduction to
Algebraic and Constructive Quantum Field Theory} (Princeton University
Press, Princeton, 1992).


\bibitem[Ba1]{B1} J. Baez, Lett. Math. Phys. {\bf 31}, 213 (1994).

\bibitem[Ba2]{B2} J. Baez, {\it Diffeomorphism Invariant Generalized
Measures on the Space of Connections Modulo Gauge Transformations},
in Proceedings of the Conference on Quantum
Topology, ed. D. Yetter (World Scientific, Singapore, 1994). 




\bibitem[GV]{GV} I. M. Gelfand, N. Vilenkin, {\it Generalized Functions},
vol. IV (Academic Press, New York, 1964).



\bibitem[GT]{GT} R. Gambini, A. Trias, Phys. Rev. {\bf D23}, 553 (1981).

\bibitem[GJ]{GJ} J. Glimm, A. Jaffe, {\it Quantum Physics}
(Springer Verlag, New York, 1987).


\bibitem[Ki]{Ki} A. A. Kirillov, {\it Elements of the Theory of 
Representations} (Springer Verlag, Berlin, 1975).


\bibitem[MM]{MM} D. Marolf, J. M. Mour\~ao, Comm. Math. Phys. 
{\bf 170}, 583 (1995).

\bibitem[MTV]{MTV} J. M. Mour\~ao, T. Thiemann, J. M. Velhinho, J. Math. Phys.
{\bf 40}, 2337 (1999).

\bibitem[Ma]{Ma} J. Martins, {\it Mec\^anica Qu\^antica em Espa\c cos de 
Conex\~oes} (Instituto Superior T\'ecnico, 2000), unpublished.

\bibitem[ReSi]{RS2} M. Reed, B. Simon, {\it Methods of Modern Mathematical
Physics II: Fourier Analysis, Self-Adjointeness} (Academic Press, 1975).


\bibitem[RoSm]{RS} C. Rovelli, L. Smolin, Nucl. Phys. {\bf B331}, 80 (1990).



\bibitem[T]{T} T. Thiemann, {\it Introduction to Modern Canonical Quantum
General Relativity}, to appear in "Living Reviews".



\bibitem[Va1]{Va1} M. Varadarajan, Phys. Rev. {\bf D61}, 104001 (2000). 

\bibitem[Va2]{Va2} M. Varadarajan, Phys. Rev. {\bf D64}, 104003 (2001). 


\bibitem[Ve2]{Ve2} J. M. Velhinho, {\it M\'etodos Matem\'aticos em 
Quantiza\c c\~ao Can\'onica de Espa\c cos de Fase n\~ao Triviais},
Ph.D. Dissertation (Universidade T\'ecnica de Lisboa,  
Instituto Superior T\'ecnico, 2001).



\bibitem[Ya]{Ya} Y. Yamasaki, {\it Measures on Infinite Dimensional
Spaces} (World Scientific, Singapore,  1985).

\end{thebibliography}
\end{document}